\documentclass[prl,aps,twocolumn,draft,tightenlines,showpacs,floatfix,superscriptaddress]{revtex4}
\usepackage{amssymb}
\usepackage{psfig}
\usepackage{colordvi}

\begin{document}

\title{Spin Liquid Phase in Anisotropic Triangular Lattice Heisenberg Model }
\author{M. Q. Weng}
\affiliation{Department of Physics and Astronomy, California State University,
Northridge, CA 91330}
\author{D. N. Sheng}
\affiliation{Department of Physics and Astronomy, California State University,
Northridge, CA 91330}
\author{Z. Y. Weng}
\affiliation{Center for Advanced Study, Tsinghua University, Beijing 100084, China}
\author{Robert J. Bursill}
\affiliation{School of Physics, University of New South Wales, Sydney, New South Wales
2052, Australia}
\date{\today}

\begin{abstract}
Based on exact diagonalization and density matrix renormalization group
(DMRG) method, we show that an anisotropic triangular lattice Heisenberg
spin model has three distinct quantum phases. In particular, a spin-liquid
phase is present in the weak interchain coupling regime, which is
characterized by an anisotropic spin structure factor with an
exponential-decay spin correlator along the weaker coupling direction,
consistent with the Cs$_{2}$CuCl$_{4}$ compounds. In the obtained phase
diagram, the spin liquid phase is found to persist up to a relatively large
critical anisotropic coupling ratio $J^{\prime }/J=0.78$, which is
stabilized by strong quantum fluctuations, with a parity symmetry distinct
from two magnetic ordered states in the stronger coupling regime.
\end{abstract}

\maketitle

Two dimensional (2D) frustrated spin systems have attracted intensive
studies as they may exhibit unconventional magnetic properties \cite%
{fazekas_74,baskaran_88, kivelson_87,senthil_01}. The isotropic spin-1/2
Heisenberg antiferromagnet (HAFM) on a triangular lattice was a candidate
for the realization of a disordered spin-liquid phase \cite{fazekas_74}, but
it turns out to exhibit a three-sublattice
antiferromagnetic-long-range-order (AFLRO) as established by analytic \cite%
{trumper_99,chubukov_94,chung_01,quant_magnet} and numerical \cite%
{leung_93,capriotti_99,quant_magnet} studies. Among various spin models, a
spin-liquid phase has been established for more geometrically frustrated
systems on the kagome lattice \cite{lecheminant_97,balents_02}, dimer models 
\cite{rokhsar_88} and models involving four spin exchange terms\cite{4ring}.
The Heisenberg models on the square lattice with third-nearest-neighbor
couplings may also have a spin-liquid ground state as revealed by recent
numerical studies based on DMRG calculation\cite{capriotti_04}.

From the experimental point of view, the HAFM on an \emph{anisotropic}
triangular lattice is particularly interesting as it is directly relevant to
the quantum magnet in the Cs$_{2}$CuCl$_{4}$ compounds \cite%
{coldea_01,coldea_02,coldea_03}, which may be described by a minimal model
at half-filling \cite{yunoki_04}: 
\begin{equation}
H=J\sum_{\langle i,j\rangle }\mathbf{S}_{i}\cdot \mathbf{S}_{j}+J^{\prime
}\sum_{\langle \langle i,j\rangle \rangle }\mathbf{S}_{i}\cdot \mathbf{S}_{j}
\label{eq:ham}
\end{equation}%
Here $\mathbf{S}_{i}$ are spin-1/2 operators, and $J,$ $J^{\prime }\geq 0$
are the nearest-neighbor couplings along the chain ($J$) and the other two
axes ($J^{\prime }$) between different chains on a triangular lattice.

Based on the variational Monte Carlo (VMC) method, a resonating valence bond
(RVB) wave function was previously proposed \cite{yunoki_04} to describe the
low-lying anisotropic spin excitation observed experimentally \cite%
{coldea_01,coldea_02,coldea_03} in these systems, which suggests a gapless
spin-liquid state. The model has also been studied by different analytic
approaches such as spin wave theory (SWT) \cite{trumper_99}, large-$S$
expansion \cite{chung_01}, as well as the series expansion \cite%
{zheng_99,zheng_05}. These works have predicted magnetic ordered states at $%
J^{\prime }\geq 0.3J\pm 0.03J$ side, while the magnetic order vanishes on
the smaller $J^{\prime }$ side suggesting a disordered phase. The recent
series expansion study by Zheng \emph{et al.} \cite{zheng_05} has further
indicated that quantum renormalizations strongly enhance the
one-dimensionality of the spectra, which implies that a more accurate
description of quantum effects is needed. Thus exact calculations with
taking into account all the quantum fluctuations are highly desirable in
order to further establish the existence of spin-liquid phase, properly
determine the quantum phase diagram as well as the nature of quantum phase
transitions.

In this Letter, we present a systematic numerical study of the magnetic
phase diagram of the HAFM model in the spatially anisotropic triangular
lattice at zero temperature by using exact diagonalization (ED) and DMRG
methods. The main results are shown in Fig. 1, where three quantum phases
are found with very distinctive magnetic structure factors. At small $%
J^{\prime }/J$, the ground state is a disordered spin liquid state, which
smoothly connects to one-dimensional (1D) decoupled spin chains in the limit
of $J^{\prime }\rightarrow 0$. The long range three-sublattice spiral N\'{e}%
el ordered phase occurs at an intermediate $J^{\prime }/J\sim 1,$ and then a
collinear N\'{e}el ordered phase appears beyond a larger $J^{\prime }/J$. In
particular, we find that the regime of the spin liquid phase extends over to
a critical value, $J_{c1}^{\prime }/J=0.78\pm 0.05$, which is significantly
larger than the one for magnetic disorder phase determined by previous
theoretical approaches \cite{trumper_99,chung_01,zheng_99}. Such a phase
boundary has been reliably identified by a vanishing first excitation
energy, coinciding with the change of the ground state symmetry. By further
performing a DMRG calculation \cite{white_93,white_98, bursill} with
periodic boundary condition (PBC)\cite{newdmrg}, we are able to establish 
an exponential decay of the equal-time spin correlation function between 
different chains,
a hallmark for the 2D spin-liquid phase \cite{senthil_00}. 

Specifically, we consider a finite size system on the torus with length
vectors $\mathbf{L}_{1}=N_{1}\mathbf{a}_{1}$ and $\mathbf{L}_{2}=N_{2}%
\mathbf{a}_{2}$ connecting identical sites (i.e., a PBC system). Here $%
\mathbf{a}_{1}$ and $\mathbf{a}_{2}$ [$\mathbf{a}_{1}=(1,0)$, $\mathbf{a}%
_{2}=(1/2,\sqrt{3}/2)$ in real space] are two primitive vectors of the
triangular lattice. The coupling strength along $\mathbf{a}_{1}$ is $J$ and
along both $\mathbf{a}_{2}$ and $\mathbf{a}_{3}=\mathbf{a}_{2}-\mathbf{a}_{1}
$ is $J^{\prime }$. The total number of sites is $N=N_{1}\times N_{2}$. The
ground state is determined by a Lanczos diagonalization of the Hamiltonian
using all symmetries \cite{leung_93,bernu_94} 
for system sizes up to $N=36$ (corresponding to a Hilbert space of a
dimension $N_{H}=63092837$). On the other hand, the DMRG method \cite%
{white_98,bursill} allows us to extend the exact calculation to larger
systems up to $N=8\times 18$ sites (i.e., 8-legs) \cite{newdmrg}. 

\begin{figure}[tbph]
\centering 
\begin{minipage}{0.495\columnwidth}
\psfig{file=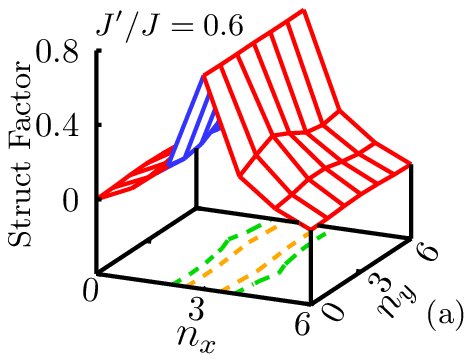,width=0.95\columnwidth}
\end{minipage}
\begin{minipage}{0.495\columnwidth}
\psfig{file=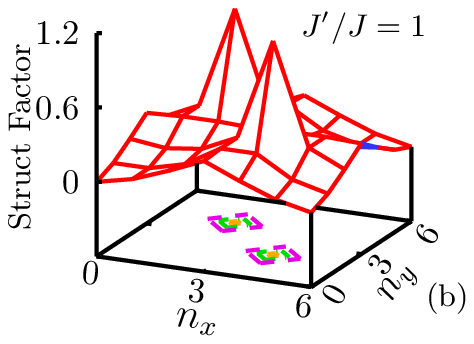,width=0.95\columnwidth}
\end{minipage}
\vskip 0.1pc 
\begin{minipage}{0.495\columnwidth}
\psfig{file=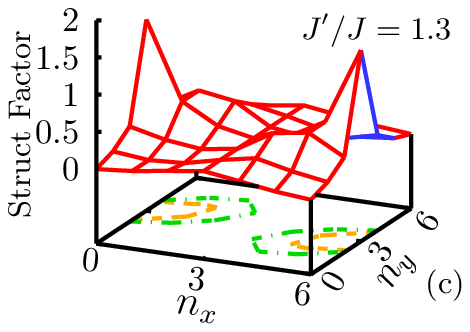,width=0.95\columnwidth}
\end{minipage}
\begin{minipage}{0.495\columnwidth}
\psfig{file=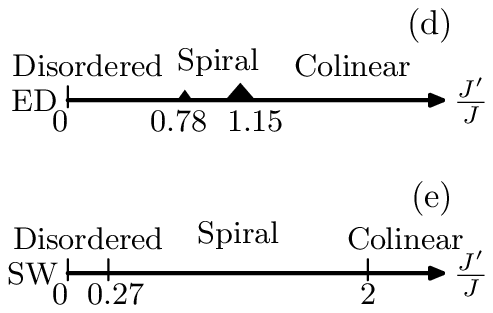,width=0.95\columnwidth}
\end{minipage}
\caption{(Color online) 
The static structure factor for $N=6\times 6$ site at different
coupling strengths: (a) $J^{\prime }/J=0.6$; (b) $J^{\prime }/J=1$; (c) $%
J^{\prime }/J=1.3$. (d) and (e) are the phase diagrams obtained from the ED
in the present work and the linear SWT\protect\cite{trumper_99},
respectively. }
\label{sf}
\end{figure}

We first present the structure factor (SF) of the statistic spin correlation 
$S(\mathbf{Q})=\sum_{ij}e^{i\mathbf{Q}\cdot (\mathbf{R}_{i}-\mathbf{R}%
_{j})}\langle S_{j}^{z}S_{i}^{z}\rangle /N$ in Fig.~\ref{sf} for a system of 
$6\times 6$ sites at $J^{\prime }=0.6J$, $J^{\prime }=J,$ and $J^{\prime
}=1.3J$, respectively. 
For the smallest $J^{\prime }$ (shown in Fig. 1(a)), $S(\mathbf{Q})$ is
peaked along the lines $n_{x}=\pm 3$, corresponding to magnetic wavevectors $%
Q_{x}=\pm \pi $, while it only shows a slight dependence on $n_{y}$ [the
components of $\mathbf{n}$ used here are related to the wavevectors $\mathbf{%
Q}$ by $Q_{x}=2\pi n_{x}/N_{1}$ and $Q_{y}=4\pi (n_{x}/2N_{1}+n_{y}/N_{2})/%
\sqrt{3}$]. This is a typical feature of weakly coupled spin-chains with
strong antiferromagnetic spin correlations within each chain, whereas the
correlations between the chains are short ranged (see also Fig. 3 below). As 
$J^{\prime }$ is increased towards the isotropic point $J^{\prime }=J$
(shown in Fig. 1(b)), $S(\mathbf{Q})$ is qualitatively changed with the
peaks moving to $\mathbf{n}=\pm (4,-2),$ corresponding to wavevectors $%
\mathbf{Q}=\pm (4\pi /3,0)$ which represents the three-sublattice (spiral) N%
\'{e}el state, in agreement with the AFLRO ground state for the isotropic
triangular lattice \cite{bernu_94,capriotti_99,chung_01}. In the region of $%
0.78<J^{\prime }<1.15$, besides the major peaks at $\mathbf{n}=\pm (4,-2)$,
there are also minor peaks at $\mathbf{n=}(0,\pm 3)$ as well, which may
indicate the incommensurate spin correlation in this region. When $J^{\prime
}$ is further increased to $J^{\prime }/J=1.3$ (shown in Fig. 1(c)), the
peaks of $S(\mathbf{Q})$ move to $\mathbf{n=}(0,\pm 3)$ or $\mathbf{Q}%
=(0,\pm 2\pi /\sqrt{3})$, which are exactly the ordering vectors for a
long-range collinear N\'{e}el order along both $\mathbf{a}_{2}$ and $\mathbf{%
a}_{3}$-axes.

Thus we have seen that when the anisotropy parameter $J^{\prime }/J$ varies
from $0$ to $2$, the ground states exhibit three distinct phases with the SF
structures dramatically different from each other. The corresponding phase
diagram is given in Fig. 1(d), where two critical points $J_{c1}^{\prime
}/J=0.78\pm 0.05$ and $J_{c2}^{\prime }/J=1.15\pm 0.1$ separate the spin
disordered phase from the left side (the small $J^{\prime }$ regime), the
spiral N\'{e}el ordered phase in the middle, and the collinear N\'{e}el
ordered phase on the right side (the large $J^{\prime }$ regime),
respectively.

\begin{figure}[htbp]
\caption{The excitation energy $\Delta E=(E_1-E_0)/N$, for system sizes $%
6\times 6$ ($\bullet$), $8\times 4$ ($\Box$), and $6\times 4$ ($\times$), in
units of $J$. }\label{gaps}\centering \psfig{file=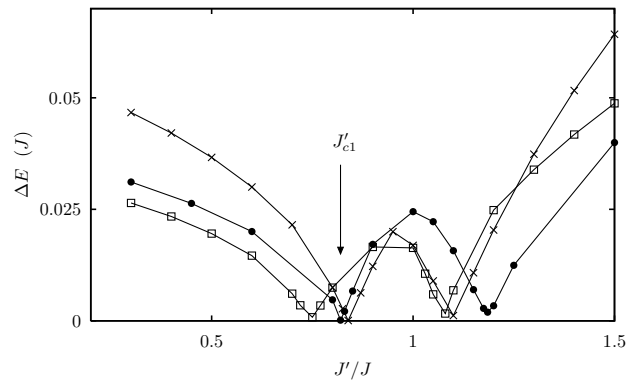,width=0.95%
\columnwidth}
\end{figure}

As $J^{\prime }/J$ crosses the phase boundaries in Fig. 1(d), not only the
characteristic SF changes, but the low energy spectrum also qualitatively
changes. In Fig.~2, $\Delta E$ as the energy difference (per site) between
the first excited state $E_{1}$ and the ground state $E_{0}$ is plotted as a
function of $J^{\prime }/J$ for three different system sizes: $N=6\times 4$, 
$8\times 4$ and $6\times 6$. As $J^{\prime }/J$ is increased from the weak
side, $\Delta E$ decreases monotonically. Remarkably, at the critical $%
J_{c1}^{\prime }$, $\Delta E$ vanishes exactly where the ground state and
the first excited state crosses each other. As a matter of a fact, the
parity symmetry (the reflection along a symmetric axis of the systems) for
the ground state also undergoes a change, from the odd parity at $J^{\prime
}<J_{c1}^{\prime }$ to the even one at $J^{\prime }>J_{c1}^{\prime }$. The
vanishing $\Delta E$ at finite $N$'s and a discrete symmetry change 
suggest a first-order phase transition occurring at $J_{c1}^{\prime }$. Here
it is noted that the qualitative behavior of the SF as a function of $%
J^{\prime }$ is independent of the system sizes in the ED calculation (with $%
N=24-36$) and the phase boundary for the spin liquid phase shown in Fig. 2
do not change much when $N$ is changed, both suggesting weak finite-size
effect in our calculation. When $J^{\prime }/J$ is further increased to
around $J_{c2}^{\prime }=1.15\pm 0.1$, $\Delta E$ reaches to an another
minimum with a finite but very small value [$\Delta E\simeq (0.0015\pm
0.0005)J]$. Together with the observation in a drastic change in the SF
around this point, it indicates that the system undergoes another phase
transition from the spiral N\'{e}el state to the collinear N\'{e}el ordered
phase. It could be a continuous transition, 
and $\Delta E$ is expected to vanish in the thermodynamic limit $%
N\rightarrow \infty $. 

\begin{figure}[tbph]
\centering \psfig{file=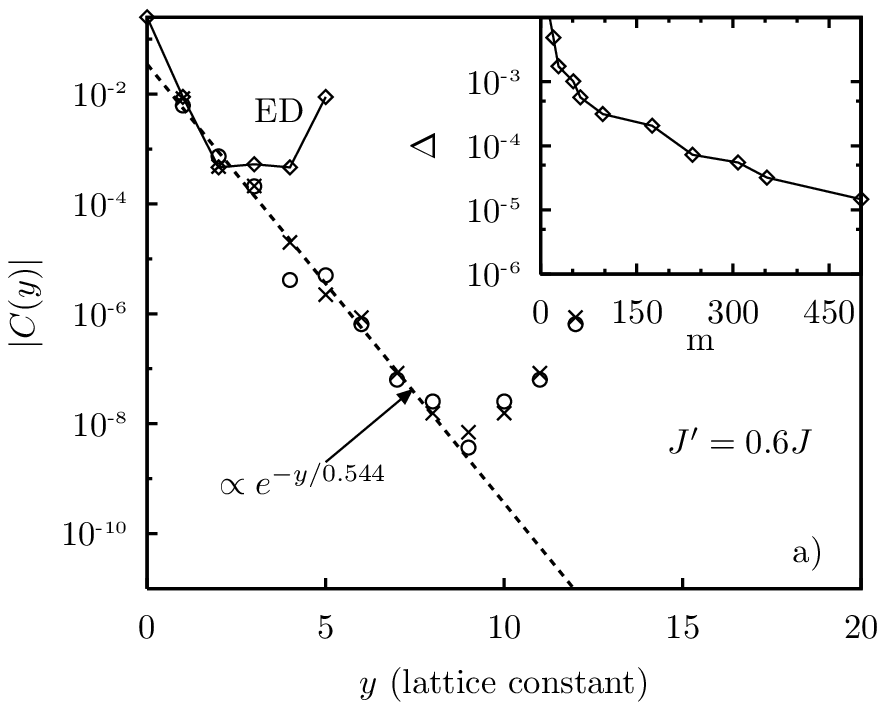,width=0.95\columnwidth} %
\psfig{file=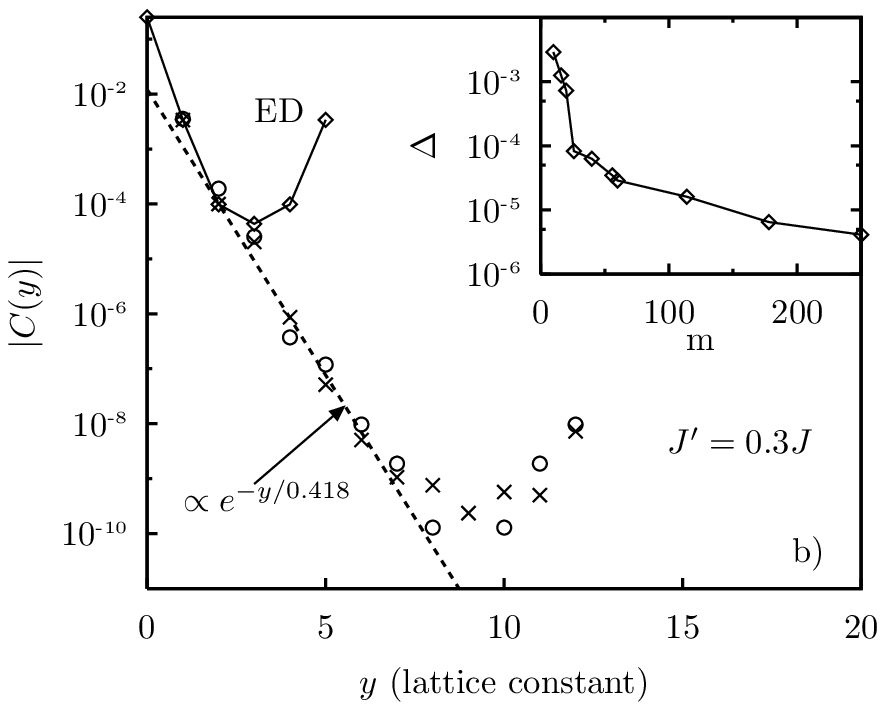,width=0.95\columnwidth}
\caption{ Absolute value of spin-spin correlation function $|C(y)|=|\langle
S_{0}^{z}S_{y\mathbf{a}_{2}}^{z}\rangle |$ along the $\mathbf{a}_{2}$
direction in six-leg system ($\times $) with $N=6\times 18$ and eight-leg
system ($\circ $) with $N=8\times 18$ at different coupling strengths: (a) $%
J^{\prime }/J=0.6$; (b) $J^{\prime }/J=0.3$. We also present the
corresponding $|C(y)|$ from the ED at $N=6\times 6$ for comparison. The
dashed lines are the fittings to the corresponding data by exponential
functions. The insets are accuracy of the ground state energy $%
\Delta=|(E(m)-E_0)/E_0|$ ($E_0$ is the exact energy from ED) as a function
of state number $m$ kept in each block in the DMRG calculation for $6\times 6
$ system. }
\label{dmrg}
\end{figure}

One of the most interesting features of the phase diagram determined by the
ED in Fig. 1(d) is that the regime 
of the spin liquid phase on the small $J^{\prime }$ side persists up to $%
J_{c1}^{\prime }\sim 0.78J$, which is much larger than $\sim 0.3J$ obtained
by the linear SWT \cite{trumper_99} shown in Fig. 1(e). 
This difference may be attributed to the importance of quantum fluctuations%
\cite{zheng_05} in the model, which could have been underestimated in the
linear SWT approach. 
We have carried out a standard SWT calculation by taking into account the
magnon scattering and the geometrical frustration \cite{leung_93,chubukov_94}
up to the second order. Our result shows that the boundary of the disordered
region indeed moves up to $J_{c1}^{\prime }=0.39J$ from the linear SWT $%
J_{c1}^{\prime }=0.27J$. The upper boundary of the spiral ordered state is
around $J^{\prime }/J\sim 1.15$, which is close to the one obtained by
series expansion in Ref.~\onlinecite{zheng_99}, while both series expansion
and spin wave theory suggest that spin disordered state may survive in a
region or at a point between spiral state and colinear state \cite%
{trumper_99}. However, we are not able to directly check this conjecture due
to the limited system size, although the fast change of SF within a small
region $1.15\leq J^{\prime }/J\leq 1.2$ might indicate the existence of a
disordered state \cite{zheng_99}.

To check the possibility of realizing an 
incommensurate spiral ordered state disfavored by PBC in small system, we
carry out the calculation with twisted boundary condition (TBC) $\mathbf{S}_{%
\mathbf{r}_{i}+\mathbf{L}_{\alpha }}=R_{\mathbf{z}}(\mathbf{Q}\cdot \mathbf{%
L_{\alpha }})\mathbf{S}_{\mathbf{r}_{i}}$ \cite{bernu_94,lecheminant_97} ($%
\alpha =1,2$), with $\mathbf{Q}$ being the classic magnetic wavevector and $%
R_{\mathbf{z}}$ being the spin-rotation operator. It is found that in the
region of $J_{c1}^{\prime }<J^{\prime }<J_{c2}^{\prime }$, the ground state
energy of $E_{0}$ with TBC is slightly lower than the one with PBC (for $%
24-36$ sites systems). Once $J^{\prime }$ goes beyond this region, PBC
always gives smaller ground state energy. 
Moreover, in the spiral ordered region ($J_{c1}^{\prime }$, $J_{c2}^{\prime }
$), the ground state energy from ED method is close to that obtained by the
series expansion $E_{0}(SE)$ \cite{zheng_99}, which may be numerically
accurate if suitable expansion states are chosen. However, outside this
region, $E_{0}$ from ED is systematically lower than the energy $E_{0}(SE)$
with spiral order, which indicates that the spiral ordered state may not be
stable outside the region of ($J_{c1}^{\prime }$,$J_{c2}^{\prime }$).

To further examine the magnetic characterization of the spin-liquid phase,
we have also performed the DMRG calculation for larger system sizes. In
particular, our DMRG code works with PBC \cite{bursill}, which can reproduce
all the ED results for $N\leq 36$ systems and extend the study up to a
system size of $8\times 18$. The system is set-up as four blocks with
replacing the single-site block in standard DMRG by a block of $N_{1}/2$
sites\cite{newdmrg}. In the insets of Fig. 3, we present the relative error
of the ground state energy $\Delta =\frac{E_{0}(m)-E_{0}}{E_{0}}$ as a
function of state number $m$ kept in the main block in the DMRG calculation
for $6\times 6$ system. Here $E_{0}$ are the ground state energy from ED
calculation. By keeping up to $m=500$, we obtain an accuracy lower than $%
\Delta =1.4\times 10^{-5}$ for PBC system. For longer length system, by
extrapolating the $E_{0}(m)$ to large $m$ limit to obtain an estimate of the
ground state energy to serve as $E_{0}$ (where no exact results exist), we
found that the error-bar remains around $10^{-5}$ for all 6-leg systems and
around $10^{-4}$ for 8-leg systems as long as we keep $m$ up to $800$ with a
discarded density matrix weight around $10^{-6}$. 
We have also applied  the standard
DMRG method\cite{white_98}  by keeping $m=1500$ states
(with adding single site to each block) and the same
ground state energies with similar error bar were obtained for both 6-leg 
and 8-leg systems.

In Fig.~\ref{dmrg}, we present the spin correlation function $C(y)=\langle
S_{0}^{z}S_{y\mathbf{a}_{2}}^{z}\rangle $ as a function of the distance
between two sites along the $\mathbf{a}_{2}$ axis (weaker coupling
direction). In Fig. 3(a) for $J^{\prime }=0.6\,J$, one first notices that
the data from the ED on a $6\times 6$ lattice agrees quite well with the
DMRG results for larger system sizes ($N=6\times 18$ and $8\times 18$). $|C({%
y})|$ decreases by about six orders of magnitude as $y$ increases, following
an exponential-decay law which can be well fitted by the dashed line: $|C({y}%
)|\propto e^{-y/0.544}$ at $y<L_{2}/2$. As our system is on a torus (PBC), $%
|C(y)|$ turns back when $y$ exceeds the half of $L_{2}$. In Fig. 3(b), $%
|C(y)|$ at a weaker $J^{\prime }=0.3\,J$ is shown, which exhibits similar
behavior, except that the exponential-decay correlation length has been
reduced from $0.544$ to $0.418$. On the other hand, along the chain
direction ($\mathbf{a}_{1})$, we have found that $C(x)$ follows a power-law
behavior at small $J^{\prime }$, consistent with the one-dimensional peaks
in the SF shown in Fig. 1(a). The remarkable similarity in the spin
correlator between $J^{\prime }=0.6\,J$ and $J^{\prime }=0.3\,J$ further
supports the ED phase diagram where spin-liquid phase extended to a large
critical $J_{c1}^{\prime }=(0.78\pm 0.05)\,J$. However we point out that we
can not exclude the possibility of the system developing an extremely small
magnetic order or breaking lattice symmetry in the thermodynamic limit,
which may not be revealed by finite-size calculations.

In conclusion, we have performed the ED and DMRG calculations of the HAFM on
an anisotropic triangular lattice. We have demonstrated that the ground
states of this system vary from the disordered spin-liquid state to a spiral
ordered state, and then to a collinear N\'{e}el ordered state, when the
anisotropic parameter $J^{\prime }/J$ is continuously increased. From the
characteristic features of the spin structure factor, the first excitation
energy, as well as the exponential decay behavior of the spin correlator, we
establish the existence and robustness of a spin-liquid phase, beyond a
previous VMC study \cite{yunoki_04}. The characterization of such a
spin-liquid phase demands further theoretical studies \cite{fisher}. Our
results provide a possible explanation of the spin-liquid behavior in the
dynamic structure factor and the quasi 1D spin excitation spectrum found in
the Cs$_{2}$CuCl$_{4}$ compounds. Stabilized by quantum fluctuations, such a
spin liquid phase is also expected to be observed experimentally in systems
with much stronger interchain couplings.

\textbf{Acknowledgment:} DNS would like to thank L. Balents for stimulating
discussions. The authors would like to thank W. Zheng for providing us the
ground state energy of series expansions. This work is supported by ACS-PRF
41752-AC10, Research Corporation grant CC5643, NSF grant/DMR-0307170
(MQW,DNS) and NSFC grants 10374058 and 90403016 (ZYW).


\begin{thebibliography}{99}
\ifx\csname natexlab\endcsname\relax
                                                                                
\fi \expandafter\ifx\csname bibnamefont\endcsname\relax
                                                                                
\fi \expandafter\ifx\csname bibfnamefont\endcsname\relax
                                                                                
\fi \expandafter\ifx\csname citenamefont\endcsname\relax
                                                                                
\fi \expandafter\ifx\csname url\endcsname\relax
                                                                                
\fi \expandafter\ifx\csname urlprefix\endcsname\relax
                                                                                
\fi \providecommand{\bibinfo}[2]{#2} \providecommand{\eprint}[2][]{\url{#2}}
                                                                                
\bibitem[Fazekas and Anderson(1974)]{fazekas_74} \bibinfo{author}{%
\bibfnamefont{P.}~\bibnamefont{Fazekas}} and \bibinfo{author}{%
\bibfnamefont{P.~W.} \bibnamefont{Anderson}}, \bibinfo{journal}{Philos. Mag.}
\textbf{\bibinfo{volume}{30}}, \bibinfo{pages}{423} (\bibinfo{year}{1974}).

\bibitem[Baskaran and Anderson(1988)]{baskaran_88} \bibinfo{author}{%
\bibfnamefont{G.}~\bibnamefont{Baskaran}} and \bibinfo{author}{%
\bibfnamefont{P.~W.} \bibnamefont{Anderson}}, \bibinfo{journal}{Phys. Rev. B}
\textbf{\bibinfo{volume}{37}}, \bibinfo{pages}{580} (\bibinfo{year}{1988}).

\bibitem[Kivelson et~al.(1987)Kivelson, Rokhsar, and Sethna]{kivelson_87} %
\bibinfo{author}{\bibfnamefont{S.~A.} \bibnamefont{Kivelson}}, %
\bibinfo{author}{\bibfnamefont{D.~S.} \bibnamefont{Rokhsar}}, and 
\bibinfo{author}{\bibfnamefont{J.~P.}
  \bibnamefont{Sethna}}, \bibinfo{journal}{Phys. Rev. B} \textbf{%
\bibinfo{volume}{35}}, \bibinfo{pages}{8865} (\bibinfo{year}{1987}).

\bibitem[Senthil and Fisher(2001{a})]{senthil_01} X. G. Wen, Phys. Rev. B 
\textbf{44}, 2664, (1999); \bibinfo{author}{\bibfnamefont{T.}~%
\bibnamefont{Senthil}} and \bibinfo{author}{\bibfnamefont{M.~P.~A.}
\bibnamefont{Fisher}}, \bibinfo{journal}{Phys. Rev.
B} \textbf{\bibinfo{volume}{63}}, \bibinfo{pages}{134521} (%
\bibinfo{year}{2001}{\natexlab{a}}); \bibinfo{journal}{Phys. Rev. Lett.}, 
\textbf{\bibinfo{volume}{86}}, \bibinfo{pages}{292} (\bibinfo{year}{2001}{%
\natexlab{b}}).

\bibitem[Schollwock et~al.(2004)Schollwock, Richter, Farnell, and Bishop]%
{quant_magnet} \bibinfo{editor}{\bibfnamefont{U.}~\bibnamefont{Schollwock}}, %
\bibinfo{editor}{\bibfnamefont{J.}~\bibnamefont{Richter}}, %
\bibinfo{editor}{\bibfnamefont{D.}~\bibnamefont{Farnell}}, and %
\bibinfo{editor}{\bibfnamefont{R.}~\bibnamefont{Bishop}}, eds., \emph{%
\bibinfo{title}{Quantum Magnetism}} (\bibinfo{publisher}{Springer}, %
\bibinfo{address}{Berlin}, \bibinfo{year}{2004}).

\bibitem[Trumper(1999)]{trumper_99} 
\bibinfo{author}{\bibfnamefont{A.~E.}
\bibnamefont{Trumper}}, \bibinfo{journal}{Phys. Rev. B} \textbf{%
\bibinfo{volume}{60}}, \bibinfo{pages}{2987} (\bibinfo{year}{1999}); J.
Merino, R. H. McKenzie, J. B. Marston, and C. H. Chung, J. Phys.: Condens.
Matter \textbf{11}, 2965 (1999); A. E. Trumper, L. Capriotti, and S.
Sorella, Phys. Rev. B, \textbf{61}, 11529 (2000).

\bibitem[Chubukov et~al.(1994)Chubukov, Sachdev, and Senthil]{chubukov_94} %
\bibinfo{author}{\bibfnamefont{A.~V.} \bibnamefont{Chubukov}}, %
\bibinfo{author}{\bibfnamefont{S.}~\bibnamefont{Sachdev}}, and %
\bibinfo{author}{\bibfnamefont{T.}~\bibnamefont{Senthil}}, %
\bibinfo{journal}{J. Phys.: Condens. Matter} \textbf{\bibinfo{volume}{6}}, %
\bibinfo{pages}{8891} (\bibinfo{year}{1994}).

\bibitem[Chung et~al.(2001)Chung, Marston, and McKenzie]{chung_01} %
\bibinfo{author}{\bibfnamefont{C.~H.} \bibnamefont{Chung}}, %
\bibinfo{author}{\bibfnamefont{J.~B.} \bibnamefont{Marston}}, and 
\bibinfo{author}{\bibfnamefont{R.~H.}
  \bibnamefont{McKenzie}}, \bibinfo{journal}{J. Phys.: Condens. Matter} 
\textbf{\bibinfo{volume}{13}}, \bibinfo{pages}{5159} (\bibinfo{year}{2001}).

\bibitem[Leung and Runge(1993)]{leung_93} \bibinfo{author}{%
\bibfnamefont{P.~W.} \bibnamefont{Leung}} and \bibinfo{author}{%
\bibfnamefont{K.~J.} \bibnamefont{Runge}}, \bibinfo{journal}{Phys. Rev. B} 
\textbf{\bibinfo{volume}{47}}, \bibinfo{pages}{5861} (\bibinfo{year}{1993}).

\bibitem{capriotti_99} L. Capriotti1, A. E. Trumper, and S. Sorella, Phys.
Rev. Lett. \textbf{82}, 3899 (1999).

\bibitem[Lecheminant et~al.(1997)Lecheminant, Bernu, Lhuillier, Pierre, and
Sindzingre]{lecheminant_97} \bibinfo{author}{\bibfnamefont{P.}~%
\bibnamefont{Lecheminant}}, \bibinfo{author}{\bibfnamefont{B.}~%
\bibnamefont{Bernu}}, \bibinfo{author}{\bibfnamefont{C.}~%
\bibnamefont{Lhuillier}}, \bibinfo{author}{\bibfnamefont{L.}~%
\bibnamefont{Pierre}}, and \bibinfo{author}{\bibfnamefont{P.}~%
\bibnamefont{Sindzingre}}, \bibinfo{journal}{Phys. Rev. B} \textbf{%
\bibinfo{volume}{56}}, \bibinfo{pages}{2521} (\bibinfo{year}{1997}).

\bibitem[Balents et~al.(2002)Balents, Fisher, and Girvin]{balents_02} %
\bibinfo{author}{\bibfnamefont{L.}~\bibnamefont{Balents}}, %
\bibinfo{author}{\bibfnamefont{M.~P.~A.} \bibnamefont{Fisher}}, and 
\bibinfo{author}{\bibfnamefont{S.~M.}
  \bibnamefont{Girvin}}, \bibinfo{journal}{Phys. Rev. B} \textbf{%
\bibinfo{volume}{65}}, \bibinfo{pages}{224412} (\bibinfo{year}{2002}); 
\bibinfo{author}{\bibfnamefont{D.~N.} \bibnamefont{Sheng}} and %
\bibinfo{author}{\bibfnamefont{L.}~\bibnamefont{Balents}}, %
\bibinfo{journal}{Phys. Rev. Lett.} \textbf{\bibinfo{volume}{94}}, %
\bibinfo{pages}{146805} (\bibinfo{year}{2005}).

\bibitem[Rokhsar and Kivelson(1988)]{rokhsar_88} \bibinfo{author}{%
\bibfnamefont{D.~S.} \bibnamefont{Rokhsar}} and \bibinfo{author}{%
\bibfnamefont{S.~A.} \bibnamefont{Kivelson}}, 
\bibinfo{journal}{Phys. Rev.
Lett.} \textbf{\bibinfo{volume}{61}}, \bibinfo{pages}{2376} (%
\bibinfo{year}{1988}); %
\bibinfo{author}{\bibfnamefont{R.}~\bibnamefont{Moessner}} and %
\bibinfo{author}{\bibfnamefont{S.}~\bibnamefont{Sondhi}}, %
\bibinfo{journal}{Phys. Rev. Lett.} \textbf{\bibinfo{volume}{86}}, %
\bibinfo{pages}{1881} (\bibinfo{year}{2001}).

\bibitem{4ring} G. Misguich, B. Bernu, C. Lhuillier, and C. Waldtmann Phys.
Rev. Lett. \textbf{81}, 1098 (1998).

\bibitem[Capriotti et~al.(2004)Capriotti, Scalapino, and White]%
{capriotti_04} \bibinfo{author}{\bibfnamefont{L.}~\bibnamefont{Capriotti}}, %
\bibinfo{author}{\bibfnamefont{D.~J.} \bibnamefont{Scalapino}}, and %
\bibinfo{author}{\bibfnamefont{S.~R.} \bibnamefont{White}}, %
\bibinfo{journal}{Phys. Rev. Lett.} \textbf{\bibinfo{volume}{93}}, %
\bibinfo{pages}{177004} (\bibinfo{year}{2004}).

\bibitem[Coldea et~al.(2001)Coldea, Tennant, Tsvelik, and Tylczynski]%
{coldea_01} \bibinfo{author}{\bibfnamefont{R.}~\bibnamefont{Coldea}}, k%
\bibinfo{author}{\bibfnamefont{D.~A.} \bibnamefont{Tennant}}, %
\bibinfo{author}{\bibfnamefont{A.~M.} \bibnamefont{Tsvelik}}, and %
\bibinfo{author}{\bibfnamefont{Z.}~\bibnamefont{Tylczynski}}, %
\bibinfo{journal}{Phys. Rev. Lett.} \textbf{\bibinfo{volume}{86}}, %
\bibinfo{pages}{1335} (\bibinfo{year}{2001}).

\bibitem[Coldea et~al.(2002)Coldea, Tennant, Habicht, Smeibidl, Wolters, and
Tylczynski]{coldea_02} \bibinfo{author}{\bibfnamefont{R.}~%
\bibnamefont{Coldea}}, 
\bibinfo{author}{\bibfnamefont{D.~A.}
\bibnamefont{Tennant}}, \bibinfo{author}{\bibfnamefont{K.}~%
\bibnamefont{Habicht}}, \bibinfo{author}{\bibfnamefont{P.}~%
\bibnamefont{Smeibidl}}, \bibinfo{author}{\bibfnamefont{C.}~%
\bibnamefont{Wolters}}, and \bibinfo{author}{\bibfnamefont{Z.}~%
\bibnamefont{Tylczynski}}, \bibinfo{journal}{Phys. Rev. Lett.} \textbf{%
\bibinfo{volume}{88}}, \bibinfo{pages}{137203} (\bibinfo{year}{2002}).

\bibitem[Coldea et~al.(2003)Coldea, Tennant, and Tylczynski]{coldea_03} %
\bibinfo{author}{\bibfnamefont{R.}~\bibnamefont{Coldea}}, %
\bibinfo{author}{\bibfnamefont{D.~A.} \bibnamefont{Tennant}}, and %
\bibinfo{author}{\bibfnamefont{Z.}~\bibnamefont{Tylczynski}}, %
\bibinfo{journal}{Phys. Rev. B} \textbf{\bibinfo{volume}{68}}, %
\bibinfo{pages}{134424} (\bibinfo{year}{2003}).


\bibitem[Yunoki and Sorella(2004)]{yunoki_04} \bibinfo{author}{%
\bibfnamefont{S.}~\bibnamefont{Yunoki}} and \bibinfo{author}{%
\bibfnamefont{S.}~\bibnamefont{Sorella}}, \bibinfo{journal}{Phys. Rev. Lett.}
\textbf{\bibinfo{volume}{92}}, \bibinfo{pages}{157003} (\bibinfo{year}{2004}%
).

\bibitem[Weihong et~al.(1999)Weihong, McKenzie, and Singh]{zheng_99} %
\bibinfo{author}{\bibfnamefont{W.}~\bibnamefont{Zheng}}, \bibinfo{author}{%
\bibfnamefont{R.~H.} \bibnamefont{McKenzie}}, and 
\bibinfo{author}{\bibfnamefont{R.~R.~P.}
  \bibnamefont{Singh}}, \bibinfo{journal}{Phys. Rev. B} \textbf{%
\bibinfo{volume}{59}}, \bibinfo{pages}{14367} (\bibinfo{year}{1999}).

\bibitem[Zheng et~al.(2005)Zheng, Fjaerestad, Singh, McKenzie, and Coldea]%
{zheng_05} \bibinfo{author}{\bibfnamefont{W.}~\bibnamefont{Zheng}}, %
\bibinfo{author}{\bibfnamefont{J.~O.} \bibnamefont{Fjaerestad}}, %
\bibinfo{author}{\bibfnamefont{R.~R.~P.} \bibnamefont{Singh}}, %
\bibinfo{author}{\bibfnamefont{R.~H.} \bibnamefont{McKenzie}}, and %
\bibinfo{author}{\bibfnamefont{R.}~\bibnamefont{Coldea}} (%
\bibinfo{year}{2005}), \bibinfo{note}{cond-mat/0506400}.

\bibitem[White(1993)]{white_93} 
\bibinfo{author}{\bibfnamefont{S.~R.}
\bibnamefont{White}}, \bibinfo{journal}{Phys. Rev. B} \textbf{%
\bibinfo{volume}{48}}, \bibinfo{pages}{10345} (\bibinfo{year}{1993}).

\bibitem[White(1998)]{white_98} 
\bibinfo{author}{\bibfnamefont{S.~R.}
\bibnamefont{White}}, \bibinfo{journal}{Phys. Rev. Lett.} \textbf{%
\bibinfo{volume}{80}}, \bibinfo{pages}{1272} (\bibinfo{year}{1998}).

\bibitem[Bursill(1999)]{bursill} 
\bibinfo{author}{\bibnamefont{R. J.
Bursill,}} \bibinfo{journal}{Phys. Rev. B} \textbf{\bibinfo{volume}{60}}, %
\bibinfo{pages}{1643} (\bibinfo{year}{1999}).

\bibitem{newdmrg} This new 2D DMRG code uses 4-blocks, each of the middle
two blocks have half of sites from one-leg (for 8-leg system with $N_1=8$,
there are $2^{N_1/2}=16$ states in this block), instead of the more
conventional 1-site structure. Thus the 2D structure as well as the PBC can
be well taken into account\cite{bursill} at the price of keeping a lot more
states in the 4-block system. We keep up to $m=800$ state in the big-block
and diagonalize 4-block system with about one third of $800\times 800 \times
2^{N_1}\sim 50\times 10^6$ (for 8-legs) states using total $S^z$ as a good
quantum number.

\bibitem[Senthil and Fisher(2000)]{senthil_00} \bibinfo{author}{%
\bibfnamefont{T.}~\bibnamefont{Senthil}} and \bibinfo{author}{%
\bibfnamefont{M.}~\bibnamefont{Fisher}}, \bibinfo{journal}{Phys. Rev. B} 
\textbf{\bibinfo{volume}{62}}, \bibinfo{pages}{7850} (\bibinfo{year}{2000}).

\bibitem[Bernu et~al.(1994)Bernu, Lecheminant, Lhuillier, and Pierre]%
{bernu_94} \bibinfo{author}{\bibfnamefont{B.}~\bibnamefont{Bernu}}, %
\bibinfo{author}{\bibfnamefont{P.}~\bibnamefont{Lecheminant}}, %
\bibinfo{author}{\bibfnamefont{C.}~\bibnamefont{Lhuillier}}, and %
\bibinfo{author}{\bibfnamefont{L.}~\bibnamefont{Pierre}}, %
\bibinfo{journal}{Phys. Rev. B} \textbf{\bibinfo{volume}{50}}, %
\bibinfo{pages}{10048} (\bibinfo{year}{1994}).

\bibitem{fisher} J. Alicea, O. I. Motrunich, and M. P. A. Fisher,
cond-mat/0508536.
\end{thebibliography}
\end{document}